\documentclass[showpacs,preprintnumbers,amsmath,amssymb]{revtex4}
\usepackage{graphicx}
\usepackage{dcolumn}
\usepackage{bm}

\begin{document}

\title{Paramagnetic Meissner Effect and Finite Spin Susceptibility\\
       in an Asymmetric Superconductor}
\author{Lianyi He, Meng Jin and Pengfei Zhuang\\
        Physics Department, Tsinghua University, Beijing 100084, China}

\begin{abstract}
A general analysis of Meissner effect and spin susceptibility of a
uniform superconductor in an asymmetric two-component fermion
system is presented in nonrelativistic field theory approach. We
found that, the pairing mechanism dominates the magnetization
property of superconductivity, and the asymmetry enhances the
paramagnetism of the system. At the turning point from BCS to
breached pairing superconductivity, the Meissner mass squared and
spin susceptibility are divergent at zero temperature. In the
breached pairing state induced by chemical potential difference
and mass difference between the two kinds of fermions, the system
goes from paramagnetism to diamagnetism, when the mass ratio of
the two species increases.
\end{abstract}
\pacs{74.20.-z, 12.38.Mh} \maketitle

\section {Introduction}
\label{s1}
It is well-known that there are two fundamental features of an
electromagnetic superconductor, the zero resistance and the
perfect diamagnetism. The latter is also called Meissner
effect\cite{fetter}. The key quantity describing the Meissner
effect is the Meissner mass or penetration depth. In the language
of gauge field theory, the Meissner mass is the mass of the
electromagnetic field obtained through the spontaneous breaking of
local $U(1)$ gauge symmetry, i.e., the Anderson-Higgs
mechanism\cite{weinberg}. Recently, the study on superconductivity
is extended to color $SU(3)$ gauge field of Quantum Chromodynamics
(QCD) at finite temperature and baryon density\cite{cscreview}.

In the linear response theory, the Meissner effect is defined in
the static and long wave limit $\omega\rightarrow 0,\vec{\bf
q}\rightarrow 0$ of the external magnetic potential $\vec{\bf A}(
\vec{\bf q})$. From the microscopic BCS theory the electric
current density can be expressed as \cite{fetter}
\begin{equation}
\label{jq} \vec{\bf j}(\vec{\bf q}) = -\vec{\bf A}(\vec{\bf
q})\frac{ne^2}{mc}\left(1+\frac{\hbar^2}{3\pi^2mn}\int_0^\infty
dpp^4\frac{\partial f(\epsilon_\Delta)}{\partial
\epsilon_\Delta}\right)\ ,
\end{equation}
where $m, e, p$ and $n$ are respectively the mass, electric
charge, momentum and density of electrons,
$\epsilon_\Delta=\sqrt{(p^2/2m-\mu)^2+\Delta^2}$ is the
quasi-particle energy with electric chemical potential $\mu$ and
energy gap $\Delta$, and $f(x)$ the fermion distribution function.
Since ${\partial f(\epsilon_\Delta)}/{\partial
\epsilon_\Delta}\leq 0$, the second term in the bracket on the
right hand side is a paramagnetic one and cancels partially the
diamagnetism characterized by the first term. However, the total
Meissner mass squared keeps positive in normal superconductor with
BCS pairing mechanism. At zero temperature, due to the limit
${\partial f(\epsilon_\Delta)}/{\partial
\epsilon_\Delta}\rightarrow -\delta(\epsilon_\Delta)$, the second
term in (\ref{jq}) disappears automatically, and there is no
paramagnetic part. In addition to the perfect diamagnetism, the
Meissner effect includes also the property of magnetic flux
expulsion upon cooling through the critical temperature
corresponding to the thermodynamic critical field.

Another quantity to describe the magnetization property of a
superconductor is the spin susceptibility. Since an electron
carries a Bohr magneton, a cold free electron gas exhibits Pauli
paramagnetism\cite{fetter}. However, at zero temperature the spin
susceptibility $\chi$ of a metallic superconductor is zero, it
does not possess Pauli paramagnetism. The physical picture is
clear: The two electrons in a Cooper pair carry opposite spin.
From the microscopic BCS theory, the spin susceptibility of a
superconductor can be written as\cite{fetter}
\begin{equation}
\frac{\chi}{\chi_P}=\frac{2}{3\pi^2}\frac{\epsilon_F}{n}\int_0^\infty
dp p^2\left(-\frac{\partial f(\epsilon_\Delta)}{\partial
\epsilon_\Delta}\right)\ ,
\end{equation}
where $\chi_P$ is the Pauli susceptibility of a normal electron
gas, and $\epsilon_F$ the electron energy at the Fermi surface.
Due to the above mentioned limit of ${\partial
f(\epsilon_\Delta)}/{\partial \epsilon_\Delta}$, the spin
susceptibility of BCS type superconductor is zero at $T=0$. At
finite temperature, $\chi$ is nonzero because of the thermo
excitation in the superconductor.

The above discussed diamagnetic Meissner mass and zero spin
susceptibility are only for normal BCS superconductor where the
two fermions participating in a Cooper pair are symmetric, i.e.,
they have the same chemical potential, the same mass, and in turn
the same Fermi surface. In many physical cases, however, the
difference in chemical potentials, or number densities, or masses
of the two kinds of fermions results in mismatched Fermi surfaces.
Such physical systems can be realized in, for instance, a
superconductor in an external magnetic field\cite{sarma} or a
strong spin-exchange field\cite{fulde,larkin,takada}, an
electronic gas with two species of electrons from different
bands\cite{liu}, a superconductor with overlapping
bands\cite{suhl,kondo}, a system of trapped ions with dipolar
interactions\cite{cirac}, a mixture of two species of fermionic
cold atoms with different densities and/or
masses\cite{liu,caldas}, an isospin asymmetric nuclear matter with
proton-neutron pairing\cite{sedrakian}, and a neutral quark matter
in dense
QCD\cite{schafer,alford,rajagopal,steiner,alford2,huang,abuki,ruster,blaschke,shovkovy,huang2,alford3}.
In the study on superconductivity in an external magnetic field,
Sarma \cite{sarma} found an interesting spatial uniform state
where there exist gapless modes. However, compared with the fully
gapped BCS state, the Sarma state is energetically unfavored and
therefore instable. A spatial non-uniform ground state where the
order parameter has crystalline structure was also proposed for
such type of superconductors by Fulde and Ferrell and Larkin and
Ovchinnikov, the so-called LOFF state. In this ground state, the
translational and the rotational symmetries of the system are
spontaneously broken. Recently, the above spatial uniform ground
state prompted new interest due to the work of Liu and
Wilczek\cite{liu}. They considered a system of two species of
fermions with a large mass difference. The stability of the state
has been discussed in many
papers\cite{wu,liu2,forbes,shovkovy,huang2}. It is now accepted
that the Sarma instability can be avoided by two possible ways,
finite difference in number densities of the two
species\cite{forbes,huang2} or a proper momentum structure of the
attractive interaction between fermions\cite{forbes}. In these
states, the dispersion relation of one branch of the
quasi-particles has two zero points at momenta $p_1$ and $p_2$,
and at these two points it needs no energy for quasi-particle
excitations. The superfluid Fermi liquid phase in the regions
$p<p_1$ and $p>p_2$ is breached by a normal Fermi liquid phase in
the region $p_1<p<p_2$. The temperature behavior of such a
breached pairing (BP) state is very different from that of a BCS
state, the temperature corresponding to the maximum gap is not
zero but finite\cite{sedrakian,liao,huang2}.

Since the dispersion relation controls the Meissner mass and the
spin susceptibility, as shown above, it is natural to guess that
the change in $\epsilon_\Delta$ in breached pairing superconductor
will modify the Meissner effect and spin susceptibility
significantly. Recently, it is found that the Meissner mass
squared of some gluons in two flavor neutral color superconductor
are negative\cite{huang3,huang4}, which indicates that the quark
matter in breached pairing state exhibits a paramagnetic Meissner
effect (PME). In condensed matter physics, PME was observed first
in high temperature superconductors such as ceramic samples of
$Bi_2Sr_2CaCu_2O_{8+\delta}$\cite{svedlindh,braunisch}, and later
in conventional superconductors such as
Nb\cite{thompson,kosti,araujo,barbara}. It was suggested that the
paramagnetic response might be a manifestation of $d$-wave
superconductivity\cite{rice}. However, it seems that it is not
necessarily to do anything with the $d$-wave analysis for the PME
in conventional superconductors\cite{kosti}. It is now widely
accepted that the PME in these materials is most likely due to
extrinsic mesoscopic or nanoscale disorder\cite{lang}. In this
paper, we will investigate the Meissner effect and spin
susceptibility in an asymmetric two-component fermion system in
nonrelativistic case, and try to prove that the Meissner effect
and spin susceptibility of a superconductor is dominated by the
pairing mechanism, and the paramagnetism and nonzero spin
susceptibility are universal phenomena of superconductors with
mismatched Fermi surfaces. We will model the pairing interaction
by a four-fermion point coupling, which is appropriate for both
electronic system, cold fermionic atom gas, nuclear matter and
dense quark matter. Since our purpose is a general analysis for
the Meissner effect and magnetization property, we will neglect
the inner structures of fermions like spin, isospin, flavor, and
color, which are important and bring much abundance while are not
central for pairing.

The paper is organized as follows. In Section \ref{s2}, we review
the BCS theory in a symmetric fermion system and show how to
calculate the Meissner mass and spin susceptibility in a
nonrelativistic field theory approach. In section \ref{s3}, we
extend the investigation to an asymmetric two-component fermion
system with mismatched Fermi surfaces and derive the universal
formula of Meissner mass squared and spin susceptibility. We then
consider two kinds of mismatched Fermi surfaces induced by
chemical potential difference and mass difference in Sections
\ref{s4} and \ref{s5}. In Section \ref{s6}, we apply our general
discussion to a relativistic system with spin structure and, as an
example, reobtain the 8th gluon Meissner mass in neutral color
superconductor. We summarize in Section \ref{s7}. We use the
natural unit of $c=\hbar=1$ through the paper.

\section {Symmetric Fermion System}
\label{s2}
In this section we review the BCS theory, the Meissner effect and
the spin susceptibility in a symmetric fermion system in a field
theory approach. We start with a system containing two species of
fermions represented by $a$ and $b$, described by the following
nonrelativistic Lagrangian density with a four-fermion
interaction,
\begin{equation}
\label{lagr} {\cal
L}=\sum_{i=a,b}\bar{\psi}_{i}(x)\left(-\frac{\partial}{\partial
\tau}+\frac{\nabla^2}{2m}+\mu\right)\psi_{i}(x)+g\bar{\psi}_{a}(
x)\bar{\psi}_{b}(x)\psi_{b}( x)\psi_{a}(x)\ ,
\end{equation}
where $\psi(x),\bar{\psi}(x)$ are fermion fields for the two
species, and the coupling constant $g$ is positive to keep the
interaction attractive. For the symmetric system, the two species
have the same mass $m$ and chemical potential $\mu$.

The key quantity to describe a thermodynamic system is the
partition function which can be defined as
\begin{equation}
Z=\int[d\psi_a][d\bar{\psi}_a][d\psi_b][d\bar{\psi}_b]e^{\int
d\tau\int d^3x{\cal L}}
\end{equation}
in the imaginary time ($\tau$) formulism of finite temperature
field theory. According to the standard BCS approach, we introduce
the order parameter $\Delta(x)$ of superconductivity phase
transition and its complex conjugate $\Delta^*(x)$,
\begin{equation}
\Delta(x)=g\big\langle\psi_{b}( x)\psi_{a}(x)\big\rangle\ ,\ \ \
\Delta^*(x)=g\big\langle\bar{\psi}_{a}(
x)\bar{\psi}_{b}(x)\big\rangle\ ,
\end{equation}
where the symbol $\langle\ \rangle$ means ensemble average. Since
we focus in this paper on uniform and isotropic superconductor, we
take the condensate to be $x$-independent and real in the
following. Introducing the Nambu-Gorkov space\cite{fetter} defined
as
\begin{equation}
\Psi=\left(\begin{array}{c} \psi_a
\\ \bar{\psi}_b\end{array}\right)\ ,\ \ \ \bar{\Psi}=\left(\begin{array}{cc}
\bar{\psi_a}& \psi_b\end{array}\right)\ ,
\end{equation}
the partition function in mean field approximation can be written
as
\begin{equation}
\label{zmf}
Z_{MF}=\int[d\Psi][d\bar{\Psi}]e^{\int d\tau\int
d^3x\left(\bar{\Psi}{\cal G}^{-1}\Psi-|\Delta|^2/g\right)}
\end{equation}
with the inverse of the mean field fermion propagator
\begin{equation}
{\cal G}^{-1}=\left(\begin{array}{cc} -\frac{\partial}{\partial
\tau}+\frac{\nabla^2}{2m}+\mu&\Delta
\\ \Delta^*&-\frac{\partial}{\partial
\tau}-\frac{\nabla^2}{2m}-\mu\end{array}\right)\ .
\end{equation}
Taking the Gaussian integration in path integral (\ref{zmf}) and
then the Fourier transformation, the thermodynamic potential of
the system can be expressed as
\begin{equation}
\label{omega}
\Omega=\frac{\Delta^2}{g}-T\sum_n\int\frac{d^3\vec{\bf
p}}{(2\pi)^3} Tr\ln {\cal G}^{-1}(i\omega_n,\vec{\bf p})
\end{equation}
in momentum space, where $\sum_n$ is the fermion frequency
summation in the imaginary time formulism. The first term is the
mean field contribution, and the second term comes from the
quasi-particle excitations with the inverse of the propagator
\begin{equation}
{\cal G}^{-1}(i\omega_n,\vec{\bf p})=\left(\begin{array}{cc}
i\omega_n-\epsilon_p&\Delta
\\ \Delta&i\omega_n+\epsilon_p\end{array}\right)
\end{equation}
in terms of momentum and frequency $\omega_n = (2n+1)\pi T$, where
$\epsilon_p=\frac{{\bf p}^2}{2m}-\mu$ is the fermion energy.

To determine the order parameter, the occupation number of
fermions, the Meissner mass, and the spin susceptibility as
functions of temperature and chemical potential, we need to know
the fermion propagator itself. Using matrix technics it can be
easily evaluated as
\begin{equation}
\label{g0} {\cal G}(i\omega_n,\vec{\bf p})=\left(\begin{array}{cc}
{\cal G}_{11}(i\omega_n,\vec{\bf p}) &{\cal
G}_{12}(i\omega_n,\vec{\bf p})
\\ {\cal G}_{21}(i\omega_n,\vec{\bf p})&{\cal G}_{22}(i\omega_n,\vec{\bf p})\end{array}\right)\ ,
\end{equation}
with the elements
\begin{equation}
\label{element} {\cal G}_{11} = {i\omega_n+\epsilon_p\over
(i\omega_n)^2-\epsilon_\Delta^2}\ ,\ \ \ {\cal G}_{22}=
{i\omega_n-\epsilon_p\over (i\omega_n)^2-\epsilon_\Delta^2}\ ,\ \
\ {\cal G}_{12}= {-\Delta\over (i\omega_n)^2-\epsilon_\Delta^2}\
,\ \ \ {\cal G}_{21}= {-\Delta\over
(i\omega_n)^2-\epsilon_\Delta^2}\ ,
\end{equation}
where $\epsilon_\Delta =\sqrt{\epsilon_p^2+\Delta^2}$ is the
quasi-particle energy. The excitation spectra $\omega_\pm({\bf
p})$ can be read directly from the poles of the fermion
propagator,
\begin{equation}
\omega_\pm(\vec{\bf p})= \pm \epsilon_\Delta\ .
\end{equation}
It is easy to see that these excitations are all gapped with
minimal excitation energy $\Delta$.

The fermion occupation numbers $n_a=\langle\psi_a^+\psi_a\rangle,
n_b=\langle\psi_b^+\psi_b\rangle$ can be either calculated from
the derivative of the thermodynamic potential with respect to the
chemical potential, or equivalently, obtained directly from the
diagonal elements of the fermion propagator matrix,
\begin{equation}
\label{num} n_a(\vec{\bf p})=T\sum_n{\cal
G}_{11}(i\omega_n,\vec{\bf p})\ ,\ \ \ n_b(\vec{\bf
p})=-T\sum_n{\cal G}_{22}(i\omega_n,\vec{\bf p})\ .
\end{equation}
After the Matsubara frequency summation, one has
\begin{equation}
n_a(\vec{\bf p})= n_b(\vec{\bf
p})=\frac{1}{2}\left(1-\frac{\epsilon_p}{\epsilon_\Delta}\right)+\frac{\epsilon_p}{\epsilon_\Delta}f(\epsilon_\Delta)\
.
\end{equation}
At zero temperature, the fermion distribution function
$f(\epsilon_\Delta)$ goes to zero, the occupation numbers are
reduced to
\begin{equation}
\label{num2} n_a(\vec{\bf p})= n_b(\vec{\bf
p})=\frac{1}{2}\left(1-\frac{\epsilon_p}{\epsilon_\Delta}\right)\
.
\end{equation}

The gap equation which determines the gap parameter $\Delta$ as a
function of $T$ and $\mu$ self-consistently can be expressed in
terms of the non-diagonal elements of the fermion propagator
matrix,
\begin{equation}
\Delta=gT\sum_n\int\frac{d^3\vec{\bf p}}{(2\pi)^3}{\cal
G}_{12}(i\omega_n,\vec{\bf p})\ ,
\end{equation}
which is equivalent to the minimum of the thermodynamic potential,
\begin{equation}
\frac{\partial \Omega}{ \partial \Delta}=0\ .
\end{equation}
After the Matsubara frequency summation, the gap equation reads
\begin{equation}
\label{gap}
\Delta(1-gI_\Delta)=0
\end{equation}
with the function
\begin{equation}
I_\Delta=\frac{1}{2}\int\frac{d^3\vec{\bf
p}}{(2\pi)^3}\frac{1-2f(\epsilon_\Delta)}{\epsilon_\Delta}\ .
\end{equation}
It is easy to see that there are two solutions of the gap equation
(\ref{gap}): One is $\Delta=0$ which describes the symmetry phase,
and the other is $\Delta\ne 0$ determined by $1-gI_\Delta=0$ which
characterizes the symmetry breaking phase.

\subsection {Meissner Effect}
We show now how to calculate the Meissner mass in terms of the
thermodynamic potential. Suppose the fermion field carries
electric charge $e$ and couples to a magnetic potential $\vec{\bf
A}$. In mean field approximation, the magnetic potential is
treated as an external and static potential, and the thermodynamic
potential $\Omega(\vec{\bf A})$ of the system can be expanded in
powers of $\vec{\bf A}$,
\begin{equation}
\Omega(\vec{\bf A})=\Omega(0)+\frac{1}{2}M_{ij}^2A_iA_j+\ldots\ ,
\end{equation}
with the coefficient
\begin{equation}
M^2_{ij}=\frac{\partial^2 \Omega(\vec{\bf A})}{
\partial A_i\partial A_j}\Big|_{\vec{\bf A}=0}\ .
\end{equation}
Since the thermodynamic potential is just the effective potential
of the field system, the coefficients $M^2_{ij}$ can be defined as
the components of the Meissner mass squared tensor. If the ground
state of the system is isotropic, one has $M^2_{ij}=0$ for $i\neq
j$ and $M^2_{11}=M^2_{22}=M^2_{33}$, and the Meissner mass squared
$M^2$ can be defined as\cite{fukushima,hong}
\begin{equation}
M^2=\frac{1}{3}\sum_{i=1}^3\frac{\partial^2 \Omega(\vec{\bf A})}{
\partial A_i\partial A_i}\Big|_{\vec{\bf A}=0}\ .
\end{equation}

In our model of four-fermion point interaction (\ref{lagr}), the
thermodynamic potential in the presence of external and static
magnetic potential $\vec{\bf A}$ can be expressed as
\begin{equation}
\label{omega2} \Omega(\vec{\bf A})
=\frac{\Delta^2}{g}-T\sum_n\int\frac{d^3\vec{\bf p}}{(2\pi)^3} Tr
\ln {\cal G}_A^{-1}(i\omega_n,\vec{\bf p})
\end{equation}
where the $\vec{\bf A}$-dependent propagator is defined as
\begin{equation}
{\cal G}_A^{-1}(i\omega_n,\vec{\bf p})=\left(\begin{array}{cc}
i\omega_n-\epsilon_p^+&\Delta
\\ \Delta&i\omega_n+\epsilon_p^-\end{array}\right)\ ,
\end{equation}
with the fermion energies $\epsilon_p^\pm={(\vec{\bf p}\pm
e\vec{\bf A})^2\over 2m}-\mu$. To extract the Meissner mass
squared, we expand the propagator ${\cal G}_A^{-1}$,
\begin{equation}
{\cal G}_A^{-1}={\cal G}^{-1}-\frac{e}{m}\vec{\bf p}\cdot\vec{\bf
A}-\frac{e^2 A^2}{2m}\tau_3\ ,
\end{equation}
and its contribution to the thermodynamic potential,
\begin{equation}
Tr \ln {\cal G}_A^{-1} = Tr \ln {\cal G}^{-1}-\frac{e}{m} \vec{\bf
p}\cdot\vec{\bf A} Tr {\cal G}-\frac{e^2}{2m} A^2 Tr \left({\cal
G}\tau_3\right)-\frac{e^2}{2m^2}(\vec{\bf p}\cdot\vec{\bf A})^2
Tr\left( {\cal G}{\cal G}\right)+\cdots
\end{equation}
in powers of $\vec{\bf A}$, where ${\cal G}(i\omega_n,p)$ is the
propagator matrix (\ref{g0}) in the absence of magnetic field.
After the momentum integration, the linear term in $\vec{\bf A}$
vanishes, and the Meissner mass squared $M^2$ can be read from the
coefficient of the quadratic term in $A^2$ of the thermodynamic
potential $\Omega(A)$,
\begin{equation}
M^2=\frac{e^2T}{m}\sum_n\int\frac{d^3\vec{\bf
p}}{(2\pi)^3}\left({\cal G}_{11}-{\cal G}_{22}\right)
+\frac{e^2T}{m^2}\sum_n\int\frac{d^3\vec{\bf
p}}{(2\pi)^3}\frac{p^2}{3}\left({\cal G}_{11}{\cal G}_{11}+{\cal
G}_{22}{\cal G}_{22}+2{\cal G}_{12}{\cal G}_{21}\right)\ .
\end{equation}
From the comparison with the fermion occupation numbers
(\ref{num}), the first term on the right hand side is proportional
to the total number density $n$,
\begin{equation}
\frac{e^2T}{m}\sum_n\int\frac{d^3\vec{\bf p}}{(2\pi)^3}\left({\cal
G}_{11}-{\cal G}_{22}\right)=\frac{ne^2}{m}
\end{equation}
with the definition
\begin{equation}
n=\int{d^3\vec{\bf p}\over (2\pi)^3}\left(n_a(\vec{\bf
p})+n_b(\vec{\bf p})\right)\ .
\end{equation}
Employing the Matsubara frequency summation $\sum_n {\cal G}{\cal
G}$ calculated in Appendix \ref{app1} for the second term, we
recover the well-known Meissner mass squared shown in text
books\cite{fetter},
\begin{equation}
\label{m22}
M^2=\frac{ne^2}{m}\left(1+\frac{1}{3\pi^2mn}\int_0^\infty
dpp^4\frac{\partial f(\epsilon_\Delta)}{\partial
\epsilon_\Delta}\right)=\frac{n_s e^2}{m}
\end{equation}
with $n_s$ defined as
\begin{equation}
n_s=n-\frac{1}{3\pi^2m}\int_0^\infty dpp^4\left(-\frac{\partial
f(\epsilon_\Delta)}{\partial \epsilon_\Delta}\right)\ .
\end{equation}
The effective density $n_s$ is positive at any temperature and
chemical potential, which means diamagnetic superconductor for any
symmetric fermion system. At low temperature limit and in the
approach to the phase transition line of superconductivity, $n_s$
behaviors as\cite{fetter}
\begin{eqnarray}
\frac{n_s}{n}&=&1-\sqrt{\frac{2\pi\Delta_0}{T}}e^{-\Delta_0/T}\
,\ \ \ \ T\rightarrow0\nonumber\\
\frac{n_s}{n} &=&2(1-\frac{T}{T_c})\ , \ \ \ T\rightarrow T_c\ ,
\end{eqnarray}
where $\Delta_0$ is the order parameter calculated by the gap
equation (\ref{gap}) at zero temperature, and $T_c$ the critical
temperature determined by $1-gI_0(T_c) = 0$. It is also necessary
to note that the Meissner mass squared (\ref{m22}) satisfies the
renormalization condition
\begin{equation}
M^2(\Delta=0)=0\ .
\end{equation}

\subsection {Spin Susceptibility}
Assuming the thermodynamic potential in the presence of a constant
magnetic field $B$ to be $\Omega(B)$, the magnetic moment ${\cal
M}$ and the spin susceptibility $\chi$ of the system are defined
as
\begin{equation}
{\cal M} =-\frac{\partial \Omega(B)}{
\partial B}\Big|_{B=0}\ ,\ \ \ \chi=-\frac{\partial^2 \Omega(B)}{
\partial B^2}\Big|_{B=0}\ .
\end{equation}
In our model, the thermodynamic potential $\Omega(B)$ in mean
field approximation can be expressed as
\begin{equation}
\Omega(B) =\frac{\Delta^2}{g}-T\sum_n\int\frac{d^3\vec{\bf
p}}{(2\pi)^3} Tr \ln {\cal G}_B^{-1}(i\omega_n,\vec{\bf p})\ ,
\end{equation}
where the $B$-dependent propagator is defined as
\begin{equation}
\label{gb} {\cal G}_B^{-1}(i\omega_n,\vec{\bf
p})=\left(\begin{array}{cc} i\omega_n-\epsilon_p^\uparrow &\Delta
\\ \Delta&i\omega_n+\epsilon_p^\downarrow\end{array}\right)
\end{equation}
with the spin-up and spin-down fermion energies
$\epsilon_p^\uparrow=\frac{p^2}{2m}-\mu+\mu_0 B$ and
$\epsilon_p^\downarrow=\frac{p^2}{2m}-\mu-\mu_0 B$, where $\mu_0$
is some elementary magneton such as the Bohr magneton or the
nucleon magneton.

To extract the magnetic moment and spin susceptibility from the
expansion of $\Omega(B)$ in powers of $B$, we take the similar way
used for the discussion of Meissner effect in the last subsection.
We expand the propagator
\begin{equation}
{\cal G}_B^{-1}={\cal G}^{-1}-\mu_0B\ ,
\end{equation}
and its contribution to the thermodynamic potential
\begin{equation}
Tr \ln {\cal G}_B^{-1} = Tr \ln {\cal G}^{-1}- \mu_0 B Tr{\cal
G}-{1\over 2}(\mu_0 B)^2 Tr\left({\cal G}{\cal G}\right)+\cdots
\end{equation}
in powers of $B$. Substituting them into the thermodynamic
potential, we obtain from the linear term in $B$ the magnetic
moment ${\cal M}$
\begin{equation}
{\cal M}=-\mu_0T\sum_n\int\frac{d^3\vec{\bf
p}}{(2\pi)^3}\left({\cal G}_{11}+{\cal
G}_{22}\right)=\mu_0(n_b-n_a)=0\ ,
\end{equation}
and from the quadratic term the spin susceptibility\cite{fetter}
\begin{equation}
\chi=-\mu_0^2T\sum_n\int\frac{d^3\vec{\bf p}}{(2\pi)^3}\left({\cal
G}_{11}{\cal G}_{11}+{\cal G}_{22}{\cal G}_{22}+2{\cal
G}_{12}{\cal G}_{21}\right)=-\frac{\mu_0^2}{2\pi^2}\int_0^\infty
dp p^2\frac{\partial f(\epsilon_\Delta)}{\partial
\epsilon_\Delta}\ ,
\end{equation}
where we have used again the Matsubara frequency summation of
$\sum_n {\cal G}{\cal G}$ shown in Appendix \ref{app1}. It is easy
to see that $\chi=0$ at $T=0$ and $\chi=\chi_P={3\mu_0^2n\over
2\epsilon_F}$ at $T=T_c$.

\section {Asymmetric Fermion System}
\label{s3}
We discuss now the fermion pairing mechanism, and the Meissner
effect and spin susceptibility in an asymmetric fermion system,
using the same approach for the symmetric system in Section
\ref{s2} . Our asymmetric two-component system with different
masses $m_a, m_b$ and different chemical potentials $\mu_a, \mu_b$
is defined through the Lagrangian density
\begin{equation}
\label{lag} {\cal
L}=\sum_{i=a,b}\bar{\psi}_{i}(x)\left[-\frac{\partial}{\partial
\tau}+\frac{\nabla^2}{2m_i}+\mu_i\right]\psi_{i}(x)+g\bar{\psi}_{a}(
x)\bar{\psi}_{b}(x)\psi_{b}( x)\psi_{a}(x)\ .
\end{equation}

The thermodynamic potential of the system in mean field
approximation is just the same as (\ref{omega}) for the symmetric
system, but the matrix elements of the fermion propagator in
Nambu-Gorkov space are different,
\begin{eqnarray}
&& \label{element} {\cal G}_{11}=
{i\omega_n-\epsilon_A+\epsilon_S\over
(i\omega_n-\epsilon_A)^2-\epsilon_\Delta^2}\ ,\ \ \ {\cal G}_{22}=
{i\omega_n-\epsilon_A-\epsilon_S\over
(i\omega_n-\epsilon_A)^2-\epsilon_\Delta^2}\ ,\nonumber\\
&& {\cal G}_{12}= {-\Delta\over
(i\omega_n-\epsilon_A)^2-\epsilon_\Delta^2}\ ,\ \ \ {\cal G}_{21}
= {-\Delta\over (i\omega_n-\epsilon_A)^2-\epsilon_\Delta^2}\ ,
\end{eqnarray}
where $\epsilon_S$ and $\epsilon_A$ are defined as
$\epsilon_S=\frac{\epsilon_p^a+\epsilon_p^b}{2}$,
$\epsilon_A=\frac{\epsilon_p^a-\epsilon_p^b}{2}$ with the fermion
energies $\epsilon_p^a={p^2\over 2m_a}-\mu_a$ and
$\epsilon_p^b={p^2\over 2m_b}-\mu_b$, and
$\epsilon_\Delta=\sqrt{\epsilon_S^2+\Delta^2}$ is the
quasi-particle energy. The dispersion relations $\omega_{\pm}(p)$
can be read from the poles of the fermion propagator,
\begin{equation}
\omega_\pm(\vec{\bf p})=\epsilon_A\pm\epsilon_\Delta\ .
\end{equation}
Without losing generality we can choose $\epsilon_A>0$ in the
following. Different from the BCS mechanism for the symmetric
fermion system, while one branch of the excitations $\omega_+$ in
the asymmetric system is always gapped, the other one $\omega_-$
can cross the momentum axis and become gapless at the momenta
$p_1$ and $p_2$, where $p_1$ and $p_2$ satisfy the equation
$\omega_-(p)=0$ and $p_F^a<p_1<p_2<p_F^b$ with $p_F^a$ and $p_F^b$
the Fermi momenta of the two species.

The phenomena of gapless excitation is directly related to the
breached pairing mechanism. The occupation numbers for fermions
$a$ and $b$ defined in (\ref{num}) become now
\begin{eqnarray}
n_a(\vec{\bf
p})&=&\frac{1}{2}\left(1-\frac{\epsilon_S}{\epsilon_\Delta}\right)f(\omega_-)
+\frac{1}{2}\left(1+\frac{\epsilon_S}{\epsilon_\Delta}\right)f(\omega_+)\ ,\nonumber\\
n_b(\vec{\bf
p})&=&-\frac{1}{2}\left(1+\frac{\epsilon_S}{\epsilon_\Delta}\right)f(\omega_-)
-\frac{1}{2}\left(1-\frac{\epsilon_S}{\epsilon_\Delta}\right)f(\omega_+)+1\
.
\end{eqnarray}
At zero temperature, they are reduced to
\begin{eqnarray}
n_a(\vec{\bf p})&=&\frac{1}{2}\left(1-\frac{\epsilon_S}{\epsilon_\Delta}\right)\theta(\epsilon_\Delta-\epsilon_A)\ ,\nonumber\\
n_b(\vec{\bf
p})&=&1-\frac{1}{2}\left(1+\frac{\epsilon_S}{\epsilon_\Delta}\right)\theta(\epsilon_\Delta-\epsilon_A)\
.
\end{eqnarray}
If there is no breached pairing, namely
$\epsilon_\Delta>\epsilon_A$, the two species have the same
occupation number
\begin{equation}
\label{num3} n_a(\vec{\bf p})= n_b(\vec{\bf
p})=\frac{1}{2}\left(1-\frac{\epsilon_S}{\epsilon_\Delta}\right)\
,
\end{equation}
which comes back to the result (\ref{num2}) for $m_a=m_b$ and
$\mu_a=\mu_b$. However, when the breached pairing happens, the
result (\ref{num3}) is valid only in the momentum regions $p<p_1$
and $p>p_2$, and in the breached pairing $p_1<p<p_2$, we have
\begin{equation}
n_a(\vec{\bf p})=0\ ,\ \ \  n_b(\vec{\bf p})=1\ .
\end{equation}
In this case, the pairing between fermions is breached by the
region $p_1<p<p_2$, the system is in normal Fermi liquid state in
the region $p_1<p<p_2$ and superconductivity state in the regions
$p<p_1$ and $p>p_2$.

For the asymmetric system, the gap parameter $\Delta$ is still
determined through the self-consistent equation (\ref{gap}), but
the function $I_\Delta$ is changed to
\begin{equation}
I_\Delta=\frac{1}{2}\int\frac{d^3\vec{\bf
p}}{(2\pi)^3}\frac{f(\omega_-)-f(\omega_+)}{\epsilon_\Delta}\ .
\end{equation}

\subsection {Meissner Effect}
Suppose the fermions $a$ and $b$ carry electric charges $eQ_a$ and
$eQ_b$ respectively, the thermodynamic potential in the presence
of a magnetic potential $\vec{\bf A}$ in mean field approximation
is still in the form of (\ref{omega2}), but the propagator matrix
is now a little bit different,
\begin{equation}
{\cal G}_A^{-1}(i\omega_n,\vec{\bf p})=\left(\begin{array}{cc}
i\omega_n-\epsilon_p^{a+}&\Delta
\\ \Delta&i\omega_n+\epsilon_p^{b-}\end{array}\right)={\cal G}^{-1}-e\Gamma_1\vec{\bf p}\cdot\vec{\bf A}-{e^2\over 2}\Gamma_2A^2
\end{equation}
with the fermion energies $\epsilon_p^{a+}={(\vec{\bf
p}+eQ_a\vec{\bf A})^2\over 2m_a}-\mu_a$,
$\epsilon_p^{b-}={(\vec{\bf p}-eQ_b\vec{\bf A})^2\over
2m_b}-\mu_b$, and matrices $\Gamma_1$ and $\Gamma_2$ defined as
\begin{equation}
\Gamma_1=\left(\begin{array}{cc} \frac{Q_a}{m_a}&0
\\ 0&\frac{Q_b}{m_b}\end{array}\right)\ ,\ \ \ \
\Gamma_2=\left(\begin{array}{cc} \frac{Q_a^2}{m_a}&0
\\ 0&-\frac{Q_b^2}{m_b}\end{array}\right)\ .
\end{equation}

\begin{widetext}
Taking again the expansion of $Tr \ln {\cal G}_A^{-1}$ in powers
of $\vec{\bf A}$,
\begin{equation}
Tr \ln {\cal G}_A^{-1}= Tr \ln {\cal G}^{-1}-e\vec{\bf
p}\cdot\vec{\bf A} Tr\left({\cal G}\Gamma_1\right)-{e^2\over 2}
A^2 Tr\left({\cal G}\Gamma_2\right)-\frac{e^2}{2}(\vec{\bf
p}\cdot\vec{\bf A})^2 Tr\left({\cal G}\Gamma_1{\cal
G}\Gamma_1\right)+\cdots\ ,
\end{equation}
the Meissner mass squared $M^2$ can be extracted from the
quadratic term in $\vec{\bf A}$ of $\Omega(\vec{\bf A})$,
\begin{eqnarray}
\label{m2}
M^2&=&M_D^2+M_P^2\ ,\nonumber\\
M_D^2 &=&e^2T\sum_n\int\frac{d^3\vec{\bf
p}}{(2\pi)^3}\left(\frac{Q_a^2}{m_a}{\cal
G}_{11}-\frac{Q_b^2}{m_b}{\cal
G}_{22}\right)\ ,\nonumber\\
M_P^2&=&e^2T\sum_n\int\frac{d^3\vec{\bf
p}}{(2\pi)^3}\frac{p^2}{3}\left(\frac{Q_a^2}{m_a^2}{\cal
G}_{11}{\cal G}_{11}+\frac{Q_b^2}{m_b^2}{\cal G}_{22}{\cal
G}_{22}+\frac{2Q_aQ_b}{m_am_b}{\cal G}_{12}{\cal G}_{21}\right)\ .
\end{eqnarray}
The diamagnetic part is related to the number densities,
\begin{equation}
M_D^2=\left(\frac{n_aQ_a^2}{m_a}+\frac{n_bQ_b^2}{m_b}\right)e^2\ ,
\end{equation}
and the paramagnetic term can, with the help of the frequency
summations in Appendix \ref{app1}, be expressed as
\begin{equation}
\label{mp2} M_P^2 = e^2\int{d^3\vec{\bf p}\over (2\pi)^3}{p^2\over
3}\Bigg[\left({Q_a\over m_a}-{Q_b\over
m_b}\right)^2u_p^2v_p^2{f(\omega_+)-f(\omega_-)\over
\epsilon_\Delta}+\left({Q_a\over m_a}u_p^2+{Q_b\over
m_b}v_p^2\right)^2f'(\omega_+)+\left({Q_a\over m_a}v_p^2+{Q_b\over
m_b}u_p^2\right)^2f'(\omega_-)\Bigg]\ ,
\end{equation}
\end{widetext}
with the definitions
\begin{equation}
u_p^2={1\over 2}\left(1+{\epsilon_S\over \epsilon_\Delta}\right)\
,\ \ \ v_p^2={1\over 2}\left(1-{\epsilon_S\over
\epsilon_\Delta}\right)=1-u_p^2\ .
\end{equation}

The first term in the square bracket of (\ref{mp2}) is a new term
resulted fully from the asymmetric property of the system. Since
$f(\omega_+)<f(\omega_-)$ at any momentum, it is always negative.
We see that the asymmetry between the paired fermions enhances the
paramagnetism of the system. The second and third terms are
negative due to the property $f'(x)<0$ for any $x$, they together
will be reduced to the paramagnetic part of $M^2$ (\ref{m22}), if
we come back to the symmetric system. At zero temperature, we have
the limit $f(x)\rightarrow \theta(-x)$ and $f'(x)\rightarrow
-\delta(x)$, the second term vanishes due to
$-\delta(\omega_+)\rightarrow 0$, and the third term disappears
only in normal superconductor but keeps negative in breached
pairing state because of the property $f'(\omega_-)\rightarrow
-\delta(\epsilon_\Delta-\epsilon_A)\neq 0$.

In the following two sections we will discuss in detail the
paramagnetism of breached pairing superconductors induced by
chemical potential difference and mass difference between the two
paired fermions. Here we just point out a singularity of the
Meissner mass squared at zero temperature in general case. At the
turning point from gapped excitation to gapless excitation where
the two roots $p_1$ and $p_2$ of $\omega_-(p)=0$ coincide,
$p_1=p_2$, the momentum integration of the third term of
(\ref{mp2}) goes to infinity, since $\int dp
\delta(\omega_-)\rightarrow \int d\omega_-\left({d\omega_-\over
dp}\right)^{-1}\delta(\omega_-)\rightarrow \infty$ due to
${d\omega_-\over dp}\rightarrow 0$ at $p_1=p_2$, and then the
total Meissner mass squared becomes negative infinity at this
point.

It is easy to check that there is still the renormalization
condition for the total Meissner mass squared,
\begin{equation}
M^2(\Delta=0)=M^2(T=T_c)=0
\end{equation}
at the critical temperature $T_c$.
\subsection {Spin Susceptibility}
As shown in Section \ref{s2}, the magnetic moment ${\cal M}$ and
spin susceptibility $\chi$ are, respectively, the coefficients of
the linear and quadratic terms in the magnetic field $B$ in the
thermodynamic potential $\Omega(B)$. For the asymmetric system,
the fermion energies in the $B$-dependent propagator matrix
(\ref{gb}) are defined as
$\epsilon_p^{a\uparrow}=\frac{p^2}{2m_a}-\mu_a+\mu_0g_aB$, and
$\epsilon_p^{b\downarrow}=\frac{p^2}{2m_b}-\mu_b-\mu_0g_b B$,
where $g_a$ and $g_b$ are constants related to the quantum numbers
of angular momentum of species $a$ and $b$.

Expanding the propagator
\begin{equation}
{\cal G}_B^{-1}={\cal G}^{-1}-\mu_0 B\Gamma_3,
\end{equation}
and
\begin{equation}
Tr \ln {\cal G}_B^{-1}= Tr \ln {\cal G}^{-1}- \mu_0 BTr({\cal
G}\Gamma_3)-{1\over 2}(\mu_0 B)^2 Tr({\cal G}\Gamma_3{\cal
G}\Gamma_3)+\cdots
\end{equation}
in powers of $B$ with the matrix
\begin{equation}
\Gamma_3=\left(\begin{array}{cc} g_a&0
\\ 0&g_b\end{array}\right)\ ,
\end{equation}
we obtain from the expansion of $\Omega(B)$ the magnetic moment
\begin{equation}
{\cal M}=-\mu_0T\sum_n\int\frac{d^3\vec{\bf
p}}{(2\pi)^3}\left(g_a{\cal G}_{11}+g_b{\cal
G}_{22}\right)=\mu_0(g_bn_b-g_an_a)\ ,
\end{equation}
\begin{widetext}
and the spin susceptibility
\begin{eqnarray}
\label{sus} \chi&=&-\mu_0^2T\sum_n\int\frac{d^3\vec{\bf
p}}{(2\pi)^3}\left(g_a^2{\cal G}_{11}{\cal G}_{11}+g_b^2{\cal
G}_{22}{\cal G}_{22}+2g_ag_b{\cal G}_{12}{\cal
G}_{21}\right)\nonumber\\
&=& -\mu_0^2\int{d^3\vec{\bf p}\over (2\pi)^3}\left[(
g_a-g_b)^2u_p^2v_p^2{f(\omega_+)-f(\omega_-)\over
\epsilon_\Delta}+\left(g_au_p^2+g_bv_p^2\right)^2f'(\omega_+)+\left(g_av_p^2+g_bu_p^2\right)^2f'(\omega_-)\right]\
,
\end{eqnarray}
\end{widetext}
where we have again taken into account the frequency summations
listed in Appendix \ref{app1}. Similar to the discussion for the
Meissner mass, the first term here is new and fully due to the
asymmetry property between the two species, and $\chi$ is
divergent at zero temperature at the turning point from gapped to
gapless excitations due to the behavior of the term with
$f'(\omega_-)$ in the limit $T\rightarrow 0$.

We now turn to the details of the Meissner effect and spin
susceptibility in breached pairing state induced by chemical
potential difference and mass difference between the two paired
fermions.

\section {Only Chemical Potential Difference}
\label{s4}
We first discuss the Meissner effect and spin susceptibility
induced by chemical potential difference only. It is convenient to
replace the chemical potentials of the two species $\mu_a$ and
$\mu_b$ by their average $\bar \mu$ and difference $\delta\mu$
defined as
\begin{equation}
\bar{\mu}=\frac{\mu_a+\mu_b}{2}\ ,\ \ \
\delta\mu=\frac{\mu_b-\mu_a}{2}\ .
\end{equation}
Without losing generality, we can set $\delta\mu>0$. With $\bar
\mu$ and $\delta\mu$, the dispersion relation of the elementary
excitations can be written as
\begin{equation}
\omega_\pm(\vec{\bf
p})=\delta\mu\pm\sqrt{\left(\frac{p^2}{2m}-\bar{\mu}\right)^2+\Delta^2}\
.
\end{equation}
It is easy to see that only under the constraint
\begin{equation}
\Delta<\delta\mu\ ,
\end{equation}
there is breached pairing in the momentum region $p_1<p<p_2$ with
\begin{equation}
p_1=\sqrt{2m\left(\bar{\mu}-\sqrt{\delta\mu^2-\Delta^2}\right)}\
,\ \ \
p_2=\sqrt{2m\left(\bar{\mu}+\sqrt{\delta\mu^2-\Delta^2}\right)}\ .
\end{equation}

\subsection {Meissner Effect}
For simplicity, we set $Q_a=Q_b=1$. In this case the diamagnetic
and paramagnetic parts of the Meissner mass squared take the form
\begin{eqnarray}
M^2_D&=&\frac{ne^2}{m}\ ,\\
M_P^2&=&\frac{e^2}{m^2}\int_0^\infty dp\frac{p^4}{6\pi^2}\left[
f^\prime(\epsilon_\Delta-\delta\mu)+
f^\prime(\epsilon_\Delta+\delta\mu)\right]\ .\nonumber
\end{eqnarray}
At zero temperature, the paramagnetic part is evaluated as
\begin{equation}
M^2_P =-\frac{e^2}{m^2}\int_0^\infty
dp\frac{p^4}{6\pi^2}\delta\left(\epsilon_\Delta-\delta\mu\right)\
.
\end{equation}
It is easy to see that only in the breached pairing state, namely,
$\Delta<\delta\mu$, $M_P^2$ is nonzero. After the momentum
integration, the total Meissner mass squared can be expressed as
\begin{equation}
M^2=\frac{ne^2}{m}\left(1-\eta\frac{\delta\mu\theta(\delta\mu-\Delta)}{\sqrt{\delta\mu^2-\Delta^2}}\right)\
,
\end{equation}
with the parameter $\eta$ defined as
\begin{equation}
\eta = \frac{p_1^3+p_2^3}{6\pi^2n}\ ,
\end{equation}
and the total fermion density
\begin{equation}
\label{den9}
n=\frac{p_1^3+p_2^3}{6\pi^2}+\frac{1}{\pi^2}\left(-\int_0^{p_1}dpp^2u_p^2+\int_{p_2}^\infty
dpp^2v_p^2\right)\ .
\end{equation}

\begin{figure}[!htb]
\begin{center}
\includegraphics[width=5cm]{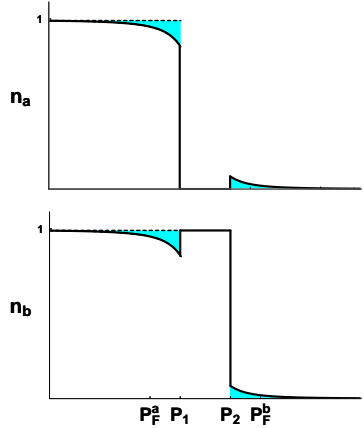}
\caption{The schematic occupation numbers $n_a$ and $n_b$ as
functions of momentum. $p_F^a$ and $p_F^b$ are Fermi momenta of
species $a$ and $b$, and $p_1$ and $p_2$ are the two roots of the
dispersion equation $\omega_-({\bf p})=0$. The pairing state is
breached by the normal fermion liquid state in the region
$p_1<p<p_2$. \label{fig1}}
\end{center}
\end{figure}

From the definition of $u_p^2$ and $v_p^2$ and their relation to
the occupation numbers $n_a(p)$ and $n_b(p)$, the first and second
integrations in (\ref{den9}) are, respectively, the upper and
lower shadow regions of $n_a(p)$ and $n_b(p)$ in three dimensional
case, shown diagrammatically in Fig.\ref{fig1}. Since
$n_a(p_1)=n_b(p_1)>1/2$ and $n_a(p_2)=n_b(p_2)<1/2$, the
contribution of the shadow regions to the total fermion density
$n$ is much smaller compared with the term $(p_1^3+p_2^3)/6\pi^2$
and we have approximately,
\begin{equation}
\label{show} \eta \simeq 1\ ,\ \ \ M^2 \simeq {ne^2\over
m}\left(1-{\delta\mu\theta(\delta\mu-\Delta)\over
\sqrt{\delta\mu^2-\Delta^2}}\right)\ ,
\end{equation}
which means global paramagnetism in the asymmetric fermion system,
if the breached pairing happens.

\subsection {Spin Susceptibility}
We take $g_a=g_b=1$ for simplicity. While the magnetic moment
${\cal M}$ disappears automatically for normal superconductor, it
is no longer zero in the breached pairing superconductor,
\begin{equation}
\label{m}
{\cal M}=\mu_0(n_b-n_a)=\mu_0\frac{p_2^3-p_1^3}{6\pi^2}\
.
\end{equation}
The physical picture is clear: while the paired fermions in the
region $p<p_1$ and $p>p_2$ have no contribution to the magnetic
moment, the unpaired fermions in the region $p_1<p<p_2$ do
contribute to ${\cal M}$.

The spin susceptibility in this case is reduced from (\ref{sus})
to
\begin{equation}
\label{chi} \chi=-\mu_0^2\int_0^\infty dp \frac{p^2}{2\pi^2}\left(
f^\prime(\epsilon_\Delta-\delta\mu)+
f^\prime(\epsilon_\Delta+\delta\mu)\right) .
\end{equation}
At zero temperature, it is nonzero only in the breached pairing
state characterized by $\Delta<\delta\mu$,
\begin{equation}
\chi=\frac{\mu_0^2m}{2\pi^2}\frac{\delta\mu}{\sqrt{\delta\mu^2-\Delta^2}}(p_1+p_2)\
.
\end{equation}
From the comparison with the Pauli susceptibility $\chi_P$, we
have the relation
\begin{equation}
{\chi\over \chi_P}\sim {\delta\mu\over
\sqrt{\delta\mu^2-\Delta^2}}\ .
\end{equation}
\section {Both Chemical Potential Difference and Mass Difference}
\label{s5}
We now consider the magnetization property of the superconductor
with both chemical potential difference and mass difference
between the paired fermions. To simplify the calculation, we still
take $Q_a=Q_b=1$ and $g_a=g_b=1$. From the dispersion relations
\begin{equation}
\omega_\pm(\vec{\bf p}) = {p^2\over
2m_A}+\delta\mu\pm\sqrt{\left({p^2\over
2m_S}-\bar\mu\right)^2+\Delta^2}
\end{equation}
with the reduced masses $m_A={2m_am_b\over m_b-m_a}$ and
$m_S={2m_am_b\over m_a+m_b}$, the condition for the system to be
in breached pairing state is
\begin{equation}
\Delta<\Delta_c=\frac{|m_b\mu_b-m_a\mu_a|}{2\sqrt{m_am_b}}={|\lambda\mu_b-\mu_a|\over
2\sqrt\lambda}\ ,
\end{equation}
where $\lambda = m_b/m_a$ is the mass ratio, and the corresponding
region of breached pairing is located at $p_1<p<p_2$ in momentum
space with
\begin{eqnarray}
\label{p12}
&&p_1=\sqrt{m_a\left[(\mu_a+\lambda\mu_b)-\sqrt{(\mu_a-\lambda\mu_b)^2-4\lambda\Delta^2}\right]}\ ,\nonumber\\
&&p_2=\sqrt{m_a\left[(\mu_a+\lambda\mu_b)+\sqrt{(\mu_a-\lambda\mu_b)^2-4\lambda\Delta^2}\right]}\
.
\end{eqnarray}

\subsection {Meissner Effect}
We now calculate analytically the Meissner mass squared at zero
temperature. When there is no breached pairing, we have from the
general expressions (\ref{m2}) to (\ref{mp2}),
\begin{equation}
M^2=\frac{(\lambda+1)ne^2}{2\lambda
m_a}-\frac{(\lambda-1)^2e^2}{\lambda^2m_a^2}\int_0^\infty dp
\frac{p^4}{6\pi^2}\frac{u_p^2v_p^2}{\epsilon_\Delta}\ .
\end{equation}
In the state with breached pairing, taking into account the
relation
\begin{equation}
{\partial\omega_-(p)\over \partial p} = p\left({u_p^2\over
m_b}-{v_p^2\over m_a}\right)={p\over \lambda
m_a}\left(u_p^2-\lambda v_p^2\right)
\end{equation}
for the integrated function with $f'(\omega_-)$ in (\ref{mp2}),
the total Meissner mass squared in the breached pairing
superconductor can be written as
\begin{widetext}
\begin{eqnarray}
\label{m2mass} M^2&=&\frac{e^2}{6\lambda m_a\pi^2}
\Bigg[p_1^3\left(\lambda-\frac{\left[1+(\lambda-1)v_1^2\right]^2}{|1-(\lambda+1)v_1^2|}\right)
+p_2^3\left(1-\frac{\left[1+(\lambda-1)v_2^2\right]^2}{|1-(\lambda+1)v_2^2|}\right)\nonumber\\
&-&3(\lambda+1)\left(\int_0^{p_1}dpp^2u_p^2-\int_{p_2}^\infty
dpp^2v_p^2\right)-2(\lambda-1)\Bigg(\int_0^{p_1}
dpp^2\frac{\epsilon_A-\delta\mu}{\epsilon_\Delta}u_p^2v_p^2+\int_{p_2}^\infty
dpp^2\frac{\epsilon_A-\delta\mu}{\epsilon_\Delta}u_p^2v_p^2\Bigg)\Bigg]
\end{eqnarray}
\end{widetext}
with the shorthand notations $v_1^2=v_{p_1}^2$ and
$v_2^2=v_{p_2}^2$.

\begin{figure}[!htb]
\begin{center}
\includegraphics[width=7cm]{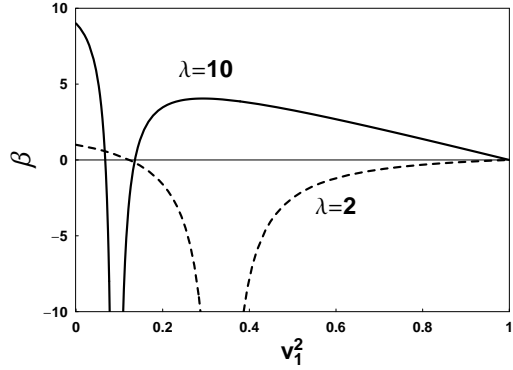}
\caption{The parameter $\beta$ as a function of $v_1^2$. The
dashed and solid lines correspond to $\lambda=2$ and $\lambda=10$,
respectively. \label{fig2}}
\end{center}
\end{figure}

It is easy to check that for $\lambda=1$, we recover the result
obtained in Section \ref{s4} for the case with only chemical
potential difference. For $\lambda>1$, we again take the
approximation of neglecting the integration terms in
(\ref{m2mass}). Considering the relation
\begin{equation}
\label{v22}
1-{\left(1+(\lambda-1)v_2^2\right)^2\over
|1-(\lambda+1)v_2^2|}\leq 0
\end{equation}
for any $\lambda$, the sign of $M^2$ depends on the quantity
\begin{equation}
\beta(\lambda,v_1^2)=\lambda-{\left(1+(\lambda-1)v_1^2\right)^2\over
|1-(\lambda+1)v_1^2|}\ .
\end{equation}
Fig.\ref{fig2} shows $\beta$ as a function of $v_1^2$ at
$\lambda=2$ and $\lambda=10$. Taking into account the condition
$v_1^2=n_a(p_1)=n_b(p_1)>1/2$, we focus on the behavior of $\beta$
in the region of $v_1^2>1/2$. At $\lambda=2$, $\beta$ is negative
in this region and results in negative Meissner mass squared and
paramagnetism of the system, which is continued with the
conclusion in Section \ref{s4}. However, at $\lambda=10$, $\beta$
becomes positive in the interesting region, the Meissner mass
squared tends to be positive and the system tends to be
diamagnetism. In fact, for very large $\lambda$, $v_2^2$ becomes
extremely small\cite{caldas}, and
$1-\left(1+(\lambda-1)v_2^2\right)^2/|1-(\lambda+1)v_2^2|\simeq
0$, the system contains approximately the diamagnetic term only.

\subsection {Spin Susceptibility}
In the case with both chemical potential difference and mass
difference, the magnetic moment ${\cal M}$ in breached pairing
state takes still the form of (\ref{m}), but $p_1$ and $p_2$
determined by the dispersion relation $\omega_-(p)=0$ are shown in
(\ref{p12}).  As for the spin susceptibility $\chi$, its general
expression at finite temperature is still (\ref{chi}). At zero
temperature, it is reduced to
\begin{equation}
\chi=\frac{\lambda
m_a\mu_0^2}{2\pi^2}\left(\frac{p_1}{|u_1^2-\lambda
v_1^2|}+\frac{p_2}{|u_2^2-\lambda v_2^2|}\right)\ .
\end{equation}

\section {Extension to Relativistic Systems}
\label{s6}
We have investigated the Meissner effect and spin susceptibility
in an asymmetric system of two kinds of fermions with different
chemical potentials, masses, charges, and magnetic moments in
nonrelativistic case. We found that the magnetization property of
breached pairing superconductor is very different from that of BCS
superconductor, and the system tends to be more paramagnetic. What
is the situation in relativistic case? Are these exotic phenomena
just a consequence of nonrelativistic kinematics? In this section,
we extend our discussion to relativistic systems, which is
relevant for the study of color superconductivity at high baryon
density\cite{huang,shovkovy,huang2,alford3,huang3,huang4}. We will
see that there is still paramagnetic Meissner effect in breached
pairing state in relativistic superconductors.
\subsection {Without Dirac Structure}
As a naive calculation, we first neglect the antiparticles and
take the same Lagrangian density (\ref{lag}). The relativistic
effect is only reflected in the fermion energies
$\epsilon_p^a=\sqrt{p^2+m_a^2}-\mu_a$,
$\epsilon_p^b=\sqrt{p^2+m_b^2}-\mu_b$. The formulas for Meissner
mass squared and spin susceptibility in Section \ref{s3} still
hold if we replace $m_i(i=a,b)$ there by $\sqrt{p^2+m_i^2}$. As an
example, we list here the diamagnetic and paramagnetic parts of
the Meissner mass squared in the case with only chemical potential
difference between the two species,
\begin{eqnarray}
M^2_D &=&{n_r e^2\over m}\ ,\\
M_P^2 &=& {e^2\over 6\pi^2}\int_0^\infty dp
\frac{p^4}{p^2+m^2}\left[f^\prime\left(\epsilon_\Delta-\delta\mu\right)+f^\prime\left(\epsilon_\Delta+\delta\mu\right)\right]\nonumber
\end{eqnarray}
with the relativistic fermion density
\begin{eqnarray}
n_r=\int_0^\infty\frac{d^3\vec{\bf
p}}{(2\pi)^3}\frac{m}{\sqrt{p^2+m^2}}\left[n_a(p)+n_b(p)\right]\ .
\end{eqnarray}
In ultra relativistic limit $m\rightarrow 0$ and at zero
temperature, they can be evaluated as
\begin{eqnarray}
M^2_D&\simeq& \frac{e^2\bar{\mu}^2}{\pi^2}\left(1+\frac{\delta
\mu^2-\Delta^2}{\bar{\mu}^2}\right)\ ,\\
M^2_P& =&
-\frac{e^2\bar{\mu}^2}{3\pi^2}\frac{\delta\mu}{\sqrt{\delta\mu^2-\Delta^2}}\left(1+\frac{\delta
\mu^2-\Delta^2}{\bar{\mu}^2}\right)\theta(\delta\mu-\Delta)\
,\nonumber
\end{eqnarray}
where in the calculation of $M_D^2$ we have taken again the
approximation used in deriving (\ref{show}). The paramagnetic part
is automatically zero in normal superconductor with $\Delta
> \delta\mu$, but negative in breached pairing superconductor with $\Delta < \delta \mu$.
Putting the two terms together, the total Meissner mass squared
can be expressed as
\begin{equation}
M^2 =  \frac{e^2\bar{\mu}^2}{3\pi^2}\left(1+\frac{\delta
\mu^2-\Delta^2}{\bar{\mu}^2}\right)\left(3-\frac{\delta\mu\theta(\delta\mu-\Delta)}{\sqrt{\delta\mu^2-\Delta^2}}\right)\
.
\end{equation}
It is negative in the case of
$\Delta<\sqrt{8/9}\delta\mu<\delta\mu$. Therefore, a relativistic
breached pairing superconductor is also paramagnetic.

\subsection {With Dirac Structure}
We study now a more realistic relativistic model containing two
kinds of fermions. The Lagrangian of the system is defined as
\begin{equation}
{\cal
L}=\sum_{\alpha=a,b}\bar{\psi}_\alpha\left(i\gamma^\mu\partial_\mu-m_\alpha+\mu_\alpha\gamma_0\right)\psi_\alpha
+g\sum_{\alpha,\beta=a,b}\left(\bar{\psi}^C_\alpha
i\gamma_5\tau_1^{\alpha\beta}\psi_\beta\right)\sum_{\alpha,\beta=a,b}\left(\bar{\psi}_{\alpha}
i\gamma_5\tau_1^{\alpha\beta}\psi^C_{\beta}\right)\ ,
\end{equation}
where $\gamma^\mu=(\gamma^0,\gamma^1,\gamma^2,\gamma^3)=(\beta,
\beta\alpha^1,\beta\alpha^2,\beta\alpha^3)$ and
$\gamma^5=i\gamma^0\gamma^1\gamma^2\gamma^3$ are Dirac matrices
with $\beta$ and $\alpha^i$ being anticommuting matrices,
$\beta^2=\alpha_i^2=1, \{\alpha_i,\alpha_j\}=0$ for $i\ne j$, and
$\{\alpha_i,\beta\}=0$, their covariant counterparts $\gamma_\mu$
and $\gamma_5$ are defined as $\gamma_\mu =
(\gamma^0,-\gamma^1,-\gamma^2,-\gamma^3)$ and $\gamma_5=\gamma^5$,
$\psi$ and $\bar{\psi}$ are Dirac spinors, $\psi^C=C\bar\psi^T$
and $\bar\psi^C=\psi^T C$ are charge-conjugate spinors,
$C=i\gamma^2\gamma^0$ is the charge conjugation matrix, the
superscript $T$ denotes transposition operation, and $\tau_1$ is
the first Pauli matrix with the elements
$\tau_1^{aa}=\tau_1^{bb}=0$ and $\tau_1^{ab}=\tau_1^{ba}=1$.

For convenience we define the Nambu-Gorkov spinors
\begin{equation}
\label{12d} \Psi
=\left(\begin{array}{c}\psi_{a}\\\psi^C_{b}\\\psi_{b}\\\psi^C_{a}\end{array}\right)\
,\ \ \ \bar\Psi
=\left(\begin{array}{cccc}\bar\psi_{a}&\bar\psi^C_{b}&\bar\psi_{b}&\bar\psi^C_{a}
\end{array}\right)\ .
\end{equation}
Introducing the order parameter
\begin{equation}
\Delta=-2g\sum_{\alpha,\beta=a,b}\Big\langle\bar{\psi}^C_\alpha
i\gamma_5\tau_{\alpha\beta}\psi_\beta\Big\rangle\ ,
\end{equation}
and taking it to be real, the thermodynamic potential of the
system in mean field approximation is still in the form of
(\ref{omega}) with the propagator matrix defined in the
4-dimensional Nambu-Gorkov space,
\begin{eqnarray}
\label{g2}
{\cal G}^{-1}=\left(\begin{array}{cccc} [{\cal
G}_0^+]_a^{-1}&i\gamma_5\Delta&0&0
\\ i\gamma_5\Delta&[{\cal G}_0^-]_b^{-1}&0&0\\ 0&0&[{\cal G}_0^+]_b^{-1}&i\gamma_5\Delta
\\ 0&0&i\gamma_5\Delta&[{\cal G}_0^-]_a^{-1}\end{array}\right)\ ,
\end{eqnarray}
where ${\cal G}_0$ is the free propagator,
\begin{eqnarray}
[{\cal G}_0^\pm]_\alpha^{-1}=i\omega_n\gamma_0-
\vec{\gamma}\cdot\vec{\bf p}-m_\alpha\pm\mu_\alpha\gamma_0\
\end{eqnarray}
with the vector matrix
$\vec{\gamma}=(\gamma^1,\gamma^2,\gamma^3)$.

Assuming that the fermion field $\psi_\alpha$ with charge $e$ can
couple to some $U(1)$ gauge potential $A_\mu$, the Meissner mass
squared can be extracted from the expansion of the thermodynamic
potential in powers of the external magnetic potential $\vec{\bf
A}$. By separating the $\vec{\bf A}$-dependent propagator
(\ref{g2}) into two terms,
\begin{equation}
{\cal G}_A^{-1}(i\omega_n,p)={\cal
G}^{-1}+e\Sigma\vec{\gamma}\cdot\vec{\bf A}\ ,
\end{equation}
with the self-energy matrix $\Sigma$ defined by
\begin{equation}
\Sigma=\left(\begin{array}{cccc} 1&0&0&0
\\ 0&-1&0&0\\ 0&0&1&0
\\ 0&0&0&-1\end{array}\right)\ ,
\end{equation}
and taking the coefficient of the quadratic term in $\vec{\bf A}$
of $\Omega(\vec{\bf A})$, the Meissner mass squared can be
evaluated as
\begin{equation}
\label{m23} M^2 =e^2T\sum_n\int\frac{d^3\vec{\bf
p}}{(2\pi)^3}\sum_{i=1}^3 Tr\Big[{\cal G}_{11}\gamma^i{\cal
G}_{11}\gamma^i+{\cal G}_{22}\gamma^i{\cal G}_{22}\gamma^i+{\cal
G}_{33}\gamma^i{\cal G}_{33}\gamma^i+{\cal G}_{44}\gamma^i{\cal
G}_{44}\gamma^i-2{\cal G}_{12}\gamma^i{\cal G}_{21}\gamma^i-2{\cal
G}_{34}\gamma^i{\cal G}_{43}\gamma^i\Big]\ .
\end{equation}

In the case of $m_a=m_b$, it is straightforward to write down
explicitly the propagator matrix elements ${\cal G}_{ij}$, by
employing the method used in \cite{huang5,he},
\begin{eqnarray}
{\cal G}_{11}(i\omega_n,p) &=&
{i\omega_n+\epsilon_p^--\delta\mu\over
(i\omega_n-\delta\mu)^2-(\epsilon_\Delta^-)^2}\Lambda_+\gamma_0+
{i\omega_n-\epsilon_p^+-\delta\mu\over (i\omega_n-\delta\mu)^2-(\epsilon_\Delta^+)^2}\Lambda_-\gamma_0\ ,\nonumber\\
{\cal G}_{22}(i\omega_n,p) &=&
{i\omega_n-\epsilon_p^--\delta\mu\over
(i\omega_n-\delta\mu)^2-(\epsilon_\Delta^-)^2}\Lambda_-\gamma_0+
{i\omega_n+\epsilon_p^+-\delta\mu\over (i\omega_n-\delta\mu)^2-(\epsilon_\Delta^+)^2}\Lambda_+\gamma_0\ ,\nonumber\\
{\cal G}_{12}(i\omega_n,p) &=& {i\Delta\over
(i\omega_n-\delta\mu)^2-(\epsilon_\Delta^-)^2}\Lambda_-\gamma_5 +
{i\Delta\over (i\omega_n-\delta\mu)^2-(\epsilon_\Delta^+)^2}\Lambda_+\gamma_5\ ,\nonumber\\
{\cal G}_{21}(i\omega_n,p) &=& {i\Delta\over
(i\omega_n-\delta\mu)^2-(\epsilon_\Delta^-)^2}\Lambda_+\gamma_5+{i\Delta\over
(i\omega_n-\delta\mu)^2-(\epsilon_\Delta^+)^2}\Lambda_-\gamma_5\ ,\nonumber\\
{\cal G}_{33}(i\omega_n,p) &=&
{i\omega_n+\epsilon_p^-+\delta\mu\over
(i\omega_n+\delta\mu)^2-(\epsilon_\Delta^-)^2}\Lambda_+\gamma_0+
{i\omega_n-\epsilon_p^++\delta\mu\over (i\omega_n+\delta\mu)^2-(\epsilon_\Delta^+)^2}\Lambda_-\gamma_0\ ,\nonumber\\
{\cal G}_{44}(i\omega_n,p) &=&
{i\omega_n-\epsilon_p^-+\delta\mu\over
(i\omega_n+\delta\mu)^2-(\epsilon_\Delta^-)^2}\Lambda_-\gamma_0+
{i\omega_n+\epsilon_p^++\delta\mu\over (i\omega_n+\delta\mu)^2-(\epsilon_\Delta^+)^2}\Lambda_+\gamma_0\ ,\nonumber\\
{\cal G}_{34}(i\omega_n,p) &=& {i\Delta\over
(i\omega_n+\delta\mu)^2-(\epsilon_\Delta^-)^2}\Lambda_-\gamma_5+
{i\Delta\over (i\omega_n+\delta\mu)^2-(\epsilon_\Delta^+)^2}\Lambda_+\gamma_5\ ,\nonumber\\
{\cal G}_{43}(i\omega_n,p) &=& {i\Delta\over
(i\omega_n+\delta\mu)^2-(\epsilon_\Delta^-)^2}\Lambda_+\gamma_5+{i\Delta\over
(i\omega_n+\delta\mu)^2-(\epsilon_\Delta^+)^2}\Lambda_-\gamma_5,
\end{eqnarray}
with the quasi-particle energies
\begin{equation}
\epsilon_p^\pm=\sqrt{p^2+m^2}\pm\bar{\mu}\ ,\ \ \
\epsilon_\Delta^\pm=\sqrt{(\epsilon_p^\pm)^2+\Delta^2}\ ,
\end{equation}
and the energy projectors
\begin{equation}
\Lambda_{\pm} = {1\over 2}\left(1\pm{\gamma_0 \vec{\gamma}\cdot
\vec{\bf p}\over \sqrt{p^2+m^2}}\right)\ .
\end{equation}

It is easy to see that in the ultra relativistic limit
$m\rightarrow 0$, the expression (\ref{m23}) is just the same as
the 8th gluon's Meissner mass squared in the two flavor gapless
color superconductor\cite{huang3,huang4}, namely,
\begin{equation}
M^2\approx
\frac{2e^2\bar{\mu}^2}{3\pi^2}\left[1-\frac{\delta\mu\theta(\delta\mu-\Delta)}{\sqrt{\delta\mu^2-\Delta^2}}\right]\
.
\end{equation}
Since we did not consider here the non-Abelian structure of the
color superconductivity, the negative Meissner mass squared for
the 8th gluon is just a reflection of the breached pairing
mechanism.

\section {Summary}
\label{s7}
We have investigated the relation between the pairing mechanism
and magnetization property of superconductivity in an asymmetric
two-component fermion system coupled to a magnetic potential. In
the frame of field theory approach, we derived the dependence of
the Meissner mass squared, magnetic moment, and spin
susceptibility of the system on the chemical potential difference,
mass difference, charge difference, and magnetic moment difference
between the two kinds of fermions. Compared with the
superconductor formed in a symmetric system where the Meissner
mass squared is globally diamagnetic, there is no magnetic moment,
and the spin susceptibility disappears at zero temperature, we
found the following new magnetization properties for the
asymmetric system with mismatched Fermi surfaces between the
paired fermions:

1) The asymmetry leads to a new paramagnetic term in the Meissner
mass squared and a new term in the spin susceptibility, and the
magnetic moment is no longer zero in asymmetric systems. Note that
the new terms and the finite magnetic moment do not depend on the
pairing mechanism, they are only the consequence of asymmetry
between the two species.

2) At the turning point from BCS to breached pairing state, the
Meissner mass squared and spin susceptibility are divergent at
zero temperature.

3) In the breached pairing state induced by chemical potential
difference between the two paired fermions, the Meissner mass
squared is paramagnetic at zero temperature.

4) In the breached pairing state with not only chemical potential
difference but also mass difference between the two kinds of
fermions, the system at zero temperature is paramagnetic at small
mass ratio and tends to be diamagnetic when the ratio is large
enough.

While the paramagnetic Meissner effect and finite spin
susceptibility discussed above are interesting, how to understand
them correctly and their reflection on physically observable
quantities are not clear. By comparing the BP and LOFF
states\cite{giannakis,giannakis2,dukelsy}, the paramagnetic
Meissner effect might be a signal of instability of the BP
state\cite{giannakis,giannakis2}. The thermodynamic potential
$\Omega(\vec{\bf Q})$ of a LOFF state with a nonzero momentum
$2\vec{\bf Q}$ of the Cooper pair can be obtained by replacing the
magnetic potential $e\vec{\bf A}$ in the thermodynamic potential
$\Omega(\vec{\bf A})$ derived above by the momentum $\vec{\bf Q}$.
If the uniform BP state is the ground state, the thermodynamic
potential must be the minimum at $\vec{\bf Q}=0$, namely,
$\partial\Omega/\partial Q_i=0$ and
$\kappa=\partial^2\Omega/\partial Q^2>0$ at $\vec{\bf Q}=0$.
Through the obvious relation $M^2=e^2\kappa$, negative Meissner
mass squared leads automatically to negative $\kappa$. Therefore,
the paramagnetic Meissner effect may be a signal that the LOFF
state is more favored than the BP state. However, the above
argument is obtained from the study for systems with fixed
chemical potentials, it is not clear if it is still true for
systems with fixed number densities of the two species. As we
know, the existence of the LOFF phase in conventional
superconductors has still not been convincingly demonstrated in
any material. If the above argument is true, the LOFF state might
be observed in trapped atomic fermion
systems\cite{mizhushima,yang}. Our research in this direction is
in progress.

{\bf Acknowledgments:}\ We thank Prof. Hai-cang Ren and Dr. Mei
Huang for stimulating discussions and comments. The work was
supported in part by the grants NSFC10428510, 10435080, 10447122
and SRFDP20040003103.

\appendix
\section{Fermion Frequency Summations}
\label{app1}
We calculate in this Appendix the fermion frequency summations
$T\sum_n{\cal G}_{11}{\cal G}_{11}$, $T\sum_n{\cal G}_{22}{\cal
G}_{22}$, and $T\sum_n{\cal G}_{12}{\cal G}_{21}$ in the Meissner
mass squared and spin susceptibility in Sections \ref{s3},
\ref{s4} and \ref{s5}. From the decomposition of these summations,
\begin{eqnarray}
T\sum_n{\cal G}_{11}{\cal
G}_{11}&=&T\sum_n\frac{(i\omega_n-\epsilon_A+\epsilon_S)^2}{[(i\omega_n-\epsilon_A)^2-\epsilon_\Delta^2]^2}
=A+(\epsilon_\Delta+\epsilon_S)^2 B+2\epsilon_S C\ ,\nonumber\\
T\sum_n{\cal G}_{22}{\cal
G}_{22}&=&T\sum_n\frac{(i\omega_n-\epsilon_A-\epsilon_S)^2}{[(i\omega_n-\epsilon_A)^2-\epsilon_\Delta^2]^2}
=A+(\epsilon_\Delta-\epsilon_S)^2 B-2\epsilon_S C\ ,\nonumber\\
T\sum_n{\cal G}_{12}{\cal
G}_{21}&=&T\sum_n\frac{\Delta^2}{[(i\omega_n-\epsilon_A)^2-\epsilon_\Delta^2]^2}=\Delta^2
B\ ,\nonumber
\end{eqnarray}
with the definitions
\begin{eqnarray}
A&=&T\sum_n\frac{1}{(i\omega_n-\epsilon_A)^2-\epsilon_\Delta^2}\
,\ \ \
B=T\sum_n\frac{1}{[(i\omega_n-\epsilon_A)^2-\epsilon_\Delta^2]^2}\
,\nonumber\\
C&=&T\sum_n\frac{1}{(i\omega_n-\epsilon_A+\epsilon_\Delta)^2(i\omega_n-\epsilon_A-\epsilon_\Delta)}\
,\nonumber
\end{eqnarray}
we need to complete the summations $A, B$ and $C$ only,
\begin{eqnarray}
A&=&\frac{f(-\omega_-)+f(\omega_+)-1}{2\epsilon_\Delta}\ ,\ \ \
B=\frac{1}{2\epsilon_\Delta}\frac{\partial}{\partial
\epsilon_\Delta}\left[\frac{f(-\omega_-)+f(\omega_+)-1}{2\epsilon_\Delta}\right]\
,\nonumber\\
C&=&\frac{f(-\omega_-)+f(\omega_+)-1}{4\epsilon_\Delta^2}-\frac{1}{2\epsilon_\Delta}\frac{\partial
f(-\omega_-)}{\partial \epsilon_\Delta}\ .\nonumber
\end{eqnarray}
Defining the function $f^{\prime}(x)=\partial f(x)/\partial x$,
we can express the summations we need as
\begin{eqnarray}
T\sum_n{\cal G}_{11}{\cal G}_{11}
&=&u_p^2v_p^2\frac{f(\omega_+)-f(\omega_-)}{\epsilon_\Delta}+v_p^4f^\prime(\omega_-)+u_p^4f^\prime(\omega_+)\ , \nonumber\\
T\sum_n{\cal G}_{22}{\cal G}_{22}
&=&u_p^2v_p^2\frac{f(\omega_+)-f(\omega_-)}{\epsilon_\Delta}+u_p^4f^\prime(\omega_-)+v_p^4f^\prime(\omega_+)\ , \nonumber\\
T\sum_n{\cal G}_{12}{\cal G}_{21}
&=&-u_p^2v_p^2\frac{f(\omega_+)-f(\omega_-)}{\epsilon_\Delta}+u_p^2v_p^2\left[f^\prime(\omega_-)+f^\prime(\omega_+)\right]\
,\nonumber
\end{eqnarray}
where the energies $\epsilon_A, \epsilon_S, \epsilon_\Delta$, the
dispersion relations $\omega_+, \omega_-$, and the functions
$u_p^2, v_p^2$ are defined in Section \ref{s3}.

\end{document}